\documentstyle[aps,eqsecnum,epsfig]{revtex}

\tighten
\begin{document}
\draft
\twocolumn[\columnwidth\textwidth\csname@twocolumnfalse\endcsname
\title{ Imaginary deuteron optical
potential due to elastic and inelastic breakup}
\date{\today}
\author{A. Ingemarsson}
\address{The Svedberg Laboratory and Department of Radiation Sciences,\\ 
 Box 533, S-75121 Uppsala, Sweden \\
E-mail address: Anders.Ingemarsson@tsl.uu.se}
\author{R. Shyam}
\address{Saha Institute of Nuclear Physics, Calcutta - 700064, India\\
E-mail address: shyam@tnp.saha.ernet.in}
\maketitle
\begin{abstract}
The contributions to the reaction cross section from the elastic  
and inelastic breakup processes, calculated within the post-form
distorted wave Born-approx\-ima\-tion theory, are used as constraints
to determine the contributions to the imaginary part of
the deuteron optical potentials (IPDOP) due to the breakup channels.
The Coulomb part of this potential due to the elastic
breakup process is seen to account for the long range absorption in the
optical potential. The nuclear parts of the IPDOP due to the 
elastic and inelastic breakup modes  
peak in different regions of the nuclear surface, with the latter
being almost an order of magnitude larger than the former. This  
makes the IPDOP due to the breakup channels determined by us 
stronger than those calculated earlier 
ignoring the inelastic breakup mode. 
 
\end{abstract}
\addvspace{5mm}] 
\vfill 
\newpage

\section{Introduction} 
 
In collisions between two nuclei the breakup of the projectile
into two or more fragments is often a strong reaction channel, which affects
not only the imaginary part but also the real part of the corresponding
optical potential. This leads to a dynamical polarization potential
(DPP) which has to be added to the real potential calculated by the
double folding (DFM) models (see eg. a recent review ~\cite{Bra97}).
Otherwise, the real part of the DFM potentials for weakly bound
projectiles (eg. $^{6,7}$Li and $^9$Be) require arbitrary re-normalization
factors in order to fit their elastic scattering data ~\cite{Sat79}. 
Whereas, folding model calculations have been
performed for both real as well as imaginary parts of the nucleon 
optical potentials~\cite{fae83}, for the case of light ions they are  
confined only to the real potentials which together with a 
phenomenological imaginary part is used to describe the corresponding 
elastic scattering.

One of the problems associated with the microscopic calculations of the 
imaginary part of the light ion optical potential has been to include
the effects due to breakup of the projectile in the field of the
target nucleus, which is a strong reaction channel for these nuclei. 
Experimental studies have shown that even for strongly bound projectiles 
the probability of breakup increases drastically with increasing beam
energy ~\cite{Wu78,Bud78}. For example, the cross section for  
breakup of the $\alpha$ particle into n + $^3$He increases by, at least,
an order of magnitude as the beam energy is varied from 65 MeV to
140 MeV~\cite{Mei83,Koo79}. Thus the effects of breakup are important
also for the tightly bound projectiles for beam energies above
30 MeV/A.

The optical potentials due to the breakup channels have been calculated by
several authors in the past ~\cite{Ian81,Ama81,Sak86,Yab92}.
Most of them are based on the coupled channels (CC)
techniques where the excitation of the breakup channel and its feedback on
the elastic channel is studied. However, such calculations are rather  
complicated as one has to find reliable approximations to include the
higher order effects and the complete breakup continuum in the
calculations ~\cite{Sak86,Yab92}. Moreover, the inelastic breakup mode,
which dominates the total breakup cross sections~\cite{Shy84} 
can not be included in these calculations. 

In this paper we follow a method introduced in ~\cite{Shy80},
where it was shown that unitarity of the scattering matrix makes it 
possible to investigate the influence of the breakup process on the
elastic scattering even without introducing the coupling
of the breakup channel back to the elastic channel. In this procedure,
the elastic scattering and breakup reaction are investigated 
separately. In the first step, the breakup of the projectile in the nuclear
and Coulomb fields of the target nucleus is calculated following the
post form distorted-wave Born-approximation (DWBA) theory.
In this first order theory, which reproduces the experimental
breakup data rather well, only the coupling of the elastic channel
to the breakup channel is considered. The contribution of
each partial wave of the incident projectile to the total breakup cross
section can be explicitly determined within this theory. Without such a 
partial wave decomposition, the present approach would have not been
feasible.

In the second step, the elastic scattering of the projectile is calculated
from the known optical potential. We determine the reaction cross section for
each partial wave (which are uniquely determined by the imaginary part of
the corresponding phase-shifts) and split it
(using the unitarity of the scattering matrix)
into two parts, one due to
the breakup channels and the another due to the rest.
Since, the reaction
cross sections out of a specific channel can be related to the
expectation value of the imaginary part of the optical potential
associated with that channel (which is calculated with the corresponding
optical model wave function in the entrance channel)~\cite{Hus85},
we use the breakup cross sections calculated within the post form DWBA theory   
as constraints in a fitting procedure to determine the imaginary 
part of the potential due to the breakup channels.
We prefer not to call it as the 
dynamical polarization potential as this phrase is used for potentials
having both real as well as imaginary parts. 

The formalism used in our calculations are discussed
in the next section. The results and their discussions are presented
in section III. The summary and conclusions of our work are given
in section IV.

\section{Method of calculations}
    
We write the phenomenologically determined optical potential $U(r)$ 
$(=V(r) + i W(r))$ as
\begin{eqnarray}
U(r) & = & [U(r)-U_{bu}(r)] + U_{bu}(r),
\end{eqnarray}
where $U_{bu}$ is the dynamical polarization potential due to the 
breakup channels. The wave functions $y_\ell(r)$ and $w_\ell(r)$, 
corresponding to potentials $U(r)$, and $[U(r)-U_{bu}(r)]$
(= $U_{bare}(r)$, the bare potential) respectively, satisfy the
following radial Schr\"odinger equations
\begin{eqnarray}
\frac{d^{2}y_{\ell}(r)}{dr^{2}} +
 [ k^{2}- U(r) -\frac{\ell(\ell+1)}{r^{2}}]
\:y_{\ell}(r) & = & 0,\\
\frac{d^{2}v_{\ell}(r)}{dr^{2}} + [ k^{2}- U_{bare}(r)
 -\frac{\ell(\ell+1)}{r^{2}}]\:w_{\ell}(r) & = & 0,
\end{eqnarray}
where $k$ is the wave-number of the incident deuteron. 
For $r > R_p$ (where $R_p$ is the distance beyond which the 
nuclear interactions can be ignored), the wave functions 
$y_{\ell}(r)$ and $w_\ell(r)$ are normalized according to
\begin{eqnarray}
y_{\ell}(r) & \sim & e^{i\delta_{\ell}}[ \cos\delta_{\ell} F_{\ell}(kr) +
\sin\delta_{\ell} G_{\ell}(kr)],\\
w_{\ell}(r) & \sim & e^{i\delta_{\ell}^0}[ \cos\delta_{\ell}^0
 F_{\ell}(kr) + \sin\delta_{\ell}^0 G_{\ell}(kr)], 
\end{eqnarray}
where $F_{\ell}$ and $G_{\ell}$ are the regular and irregular Coulomb 
functions. $\delta_{\ell}$ and $\delta_{\ell}^0$ are
the scattering phase-shifts corresponding to potentials
$U(r)$ and $U_{bare}(r)$.
 
The expressions for the partial wave amplitudes can be written in either
of the following two forms~\cite{RodTha}, 
\begin{eqnarray}
f_{\ell} & = & -\frac{1}{k}\int_{0}^{\infty}F_{\ell}(kr)U_{bare}(r)
             y_{\ell}(r) dr \nonumber \\
        & & -  \frac{1}{k}\int_{0}^{\infty}F_{\ell}(kr)U_{bu}(r)
             y_{\ell}(r) dr \nonumber \\
      & = & f_{\ell}^{A} + f_{\ell}^{B},
\end{eqnarray}
and  
\begin{eqnarray}
f_{\ell} & = & -\frac{1}{k}\int_{0}^{\infty}F_{\ell}(kr)U_{bare}(r)
            w_{\ell}(r) dr \nonumber \\
        & & -\frac{1}{k}\int_{0}^{\infty}w_{\ell}(kr)U_{bu}(r)
            y_{\ell}(r) dr \nonumber \\
      & = & f_{\ell}^{C} + f_{\ell}^{D}.
\end{eqnarray}

These decompositions of the partial wave amplitudes were used in an
earlier study~\cite{Ing96}, of the optical potential due to breakup
channels. However, the breakup amplitude was assumed to be obtained
from the difference $f_{\ell}-f_{\ell}^{C}$ instead of
$f_{\ell}-f_{\ell}^{A}$. This neglected the fact that the amplitude
from the bare potential is affected by the presence of the
breakup potential and generated a dependence on the 
real part of the breakup potential. As will be shown below, a separation
of the reaction cross section into contribution of various channels
requires the knowledge of the real potential $V(r)$ only.

Although the reaction cross section
(which includes contributions from all the inelastic channels) 
can be calculated directly from the
partial wave amplitudes described above, we use here an 
expression where it is written in terms of the imaginary part
of the optical potential~\cite{Hus85,Uda85}. This method was
used earlier~\cite{Hus84,Uda85,Chr91} to calculate the contributions 
to the imaginary part of the optical potential from the fusion channels.
We can write the reaction cross section ($\sigma_R$) as, 
\begin{eqnarray}
\sigma_R & = & \frac{2\pi}{\hbar v}<\chi_i^{(+)}|W|\chi_i^{(+)}>,
\end{eqnarray}
where $v$ is the relative velocity in the entrance channel and
$\chi_i^{(+)}$ the full solution of the Schr\"odinger equation 
(whose radial part is $y_\ell$). We can also write
\begin{eqnarray}
\sigma_R = \frac{\pi}{k^2}\sum_{\ell}(2\ell+1)T_\ell,
\end{eqnarray}
where the transmission coefficient ($T_\ell$) is given by
\begin{eqnarray}
T_{\ell} & = & \frac{4}{\hbar v} 
\int_{0}^{\infty} | y_{\ell}(r) |^{2} W(r) dr
\end{eqnarray}
It may be noted that $T_\ell$ can also 
be related to the amplitudes $f_l$ (Eqs. (6) and (7)) by
$T_\ell\, = \, 4(|f_\ell|^2 - f_\ell^I)$, where $f_\ell^I$ denotes
the imaginary part of $f_\ell$.
However, the advantage of Eq. (10) lies in
the fact that it involves a linear dependence of $T_\ell$ 
on the imaginary potential. This allows us to split
$T_\ell$ into terms corresponding to the contributions from different
channels, as will be discussed below.

Using unitarity of the $S$-matrix, the transmission coefficient
$T_{\ell}$ can be written as
\begin{equation}
T_{\ell} = 1-|S_{\ell \ell}|^2 = \sum_{c \neq \ell}|S_{\ell c }|^2,
\end{equation}
where $S$ represents the scattering matrix and $\ell$ denotes the
elastic channel. For simplicity of notation we take the projectile and
target nuclei to be spin less; hence $\ell$ corresponds to the total spin,
and c describes any other channel with total angular momentum $\ell$.
Thus Eq. (11) enables us to express the transmission
coefficient  $T_{\ell}$ as a sum (or integral for continuous channels)
over all the reaction channels. This allows us to write
\begin{eqnarray}
\sigma_R & = & \sigma_R^{bare} + \sigma_{b-up,d},
\end{eqnarray}
for each partial wave $\ell$. In this equation $\sigma_{b-up,d}$
represents the contribution to the reaction cross section from  
the breakup channels, while $\sigma_R^{bare}$ is the reaction cross
section corresponding to the remaining channels. Following 
Refs.~\cite{Hus84,Chr91}, we decompose the total imaginary potential
$W(r)$ into a bare component and a component due to breakup as,
$W(r)$ = $W_{bare}(r)$ + $W_{bu}(r)$. Then,
expressions similar to Eq. (9) can be written for 
$\sigma_R^{bare}$ and $\sigma_{b-up,d}$ with corresponding
$T_\ell$'s given by,
\begin{eqnarray}
T_{\ell}^{bare} =  \frac{4}{\hbar v} 
\int_{0}^{\infty} | y_{\ell}(r) |^{2} W_{bare}(r) dr
\end{eqnarray}
and
\begin{eqnarray}
T_{\ell}^{bu} =  \frac{4}{\hbar v} 
\int_{0}^{\infty} | y_{\ell}(r) |^{2} W_{bu}(r) dr,
\end{eqnarray}
where the $W_{bu}$ consists of a part due to the elastic breakup
$W_{diss}$ and a part due to the inelastic breakup $W_{inbu}$.
In our fitting procedure, potentials $W_{diss}$ and $W_{inbu}$
(with a certain ${\it a \; priori}$ assumed form) are varied 
so that the elastic or inelastic breakup cross sections
(calculated within the post form DWBA theory which is described below)
are reproduced for each partial wave.    
We impose the constraint that $W_{bu} \leq W $ for all $r$.
Of course, the potentials due to breakup so determined,
are specific to our breakup cross section
and it could be different from such potentials defined by other authors.

It may be noted that  dependence on the real potential 
entering into the transmission coefficients (Eqs. (10), (13) and (14))
is only through that of the phenomenological optical
potential that is used to calculate $y_\ell$. In the calculations
presented in this paper we assume that this potential is known from
the description of the elastic scattering. Thus no information 
about the real part of the potential due to the breakup channels 
can be extracted. However, had one started from
a real potential calculated within a double folding model, it would 
have been necessary to include a dynamical polarization potential 
(having a real part) in order to reproduce both the elastic scattering
and breakup probabilities simultaneously. It is also worthwhile to
note that if the 
energy dependence of the imaginary potential is known over a 
sufficiently large range of energies, the dispersion relations 
may be helpful in getting the corresponding real
potential~\cite{Nag85,Sat86}. 

In the post form DWBA theory of the inclusive breakup reaction
(eg. $d + A \rightarrow p + X$, to be represented as $(d,p)$)
the total breakup cross section is defined 
by ~\cite{Shy84,Shy80}
\begin{eqnarray}
\sigma_{b-up(d,p)} & = & \int d\Omega_p dE_p \frac{d^2 \sigma(d,p)}
                                 {d\Omega_p dE_p},
\end{eqnarray}
where $\frac{d^2\sigma(d,p)}{d\Omega_p dE_p}$ is the double differential 
cross section for the reaction $(d,p)$, which is the sum
of the elastic and inelastic breakup modes. The former (where $X$ corresponds
to $n+A(g.s.)$) is given by
\begin{eqnarray}
\frac{d^2\sigma(elastic)}{d\Omega_p dE_p} & = & \rho(phase)
                              \sum_{\ell_n m_n}\mid \beta_{\ell_n m_n}
                              \mid^2.
\end{eqnarray}
Using a zero range approximation, the amplitude $\beta_{\ell_n m_n}$ can
be written as 
\begin{eqnarray}
\beta_{\ell_n m_n} & = & D_0 \int d^3 r \chi_{p}^{(-)*}({\bf{k_p}}, 
                              \frac{A}{A+1}{\bf{r}})
                     F_{\ell_n}(k_n,r) \nonumber \\
                  & \times &  Y_{\ell_n m_n}^*(\hat r)  
                    \chi_{d}^{(+)}( {\bf{k_d,r}}) \Lambda(r) P(r).
\end{eqnarray}
$D_0$ is the zero range constant for the
$d \rightarrow p + n $ vertex. Its value has been taken to be
125 MeV fm$^{3/2}$ which is consistent with the known properties of
the p-n system. The function 
$\Lambda(r)$ takes into account the finite range effects within
the local energy approximation (LEA)~\cite{But64,Sat83}. We have used
the form of $\Lambda(r)$ as that given in~\cite{But64} with a 
finite range correction parameter of 0.621, a value used in
most of the calculations on deuteron induced transfer and breakup
reactions. $P(r)$ accounts for the nonlocality of the optical  
potentials. This is calculated by following the method of 
Perey and Buck~\cite{Per62,Sat83}, with the nonlocality parameters of 
0.85 in the neutron and proton channels and 0.54 in the deuteron
channel.
  
In Eq. (17), $\chi^{\pm}$ are the optical model wave functions
in the respective channels with $k$'s being the corresponding wave numbers.
$F_{\ell_n}(k_n,r)$ is the radial part of the wave function in the
$n + A(g.s.)$ channel.
In Eq. (16), $\rho(phase)$ is the three-body phase-space
factor~\cite{Ohls65,Fuch82}. It should be noted that the integrand
in Eq. (17) involves three scattering wave functions which are 
asymptotically oscillatory. This makes the radial integrals involved
therein very slowly converging. However, integrals of this kind
can be effectively evaluated by using a contour integration method
\cite{Vinc70,Davi88}.
                
To calculated the cross section for the inelastic breakup process, 
where $X$ can be any two-body channel of 
the $B = n + A$ system, we start from a $T$ matrix,
\begin{eqnarray}
T_{d,pX} & = & <\Phi_{BX}^{(-)}\chi_p^{(-)}|V_{np}|\phi_A \phi_d
                 \chi_d^{(+)}>,
\end{eqnarray}
where $\phi_A$ and $\phi_d$ denote the ground state wave functions of
the target nucleus A and the projectile (deuteron) respectively.
$\Phi_{BX}^{(-)}$ represents the complete scattering state of the
system B with the boundary condition X. The integration over the
internal coordinates of $\phi_A$ in Eq. (18) leads to a form factor
for the inelastic process. The calculation
of the form factor 
simplifies greatly if we use a surface approximation~\cite{Pamp78},
where we assume that the main contribution to $T_{d,pX}$ comes 
from the region outside the range of the 
nuclear interaction. The validity of this 
approximation has been tested by Kasano and Ichimura~\cite{Kasa82}
by evaluating 
this integral without recourse to this approximation. These authors
find that the surface approximation is valid for the deuteron 
induced breakup reaction even at the lower beam energy of 25 MeV. 
Thus we can represent the radial part of the form factor $(F_{\ell_n}^X)$
by its asymptotic form
\begin{eqnarray}
F_{\ell_n}^X & = &\delta_{\ell_nX}j_{\ell_n} (k_nr)+ \frac{1}{2} \sqrt{
                  \frac{m_nk_n}{m_Xk_X}} \nonumber \\
             & \times &  (S_{\ell_nX}-\delta_{\ell_nX})
                  h_{\ell_n}^{(+)}(k_nr),
\end{eqnarray}
where $j_{\ell_n}$ and $h_{\ell_n}^{(+)}$ denote the spherical Bessel 
and Hankel functions respectively. $S_{\ell_n \ell_n}$ are the 
scattering matrix elements for the elastic channel corresponding to the
angular momentum $\ell_n$. Now, it is straight forward to
carry out the integrations over the angles of the unobserved particle
to get the double differential cross section from the triple differential
cross sections. This leads to a reduced $T$ matrix for the process
$d + A \rightarrow p + X$,
\begin{eqnarray}
{\tilde T}_{d,pX} & = & \sqrt{\frac{m_nk_n}{m_Xk_X}} \frac{S_{\ell_nX}}
                       {S_{\ell_n \ell_n}-1}\nonumber \\ 
                  &  \times & D_0 \int d^3 r \chi_{p}^{(-)*}({\bf{k_p}}, 
                        \frac{A}{A+1}{\bf{r}}) \nonumber \\ 
                  & \times &  [F_{\ell_n}(k_n,r)-j_{\ell_n}(k_nr)]
                      Y_{\ell_n m_n}^*(\hat r) \nonumber \\ 
                  &  \times & \chi_{d}^{(+)}( {\bf{k_d,r}}) \Lambda(r) P(r).
\end{eqnarray}
In order to calculate the double differential cross section for the
inelastic breakup, one has to
sum over all the channels $X \neq \ell_n$. Since, in Eq.(20), the entire
dependence on channel $X$ rests solely in the $S$ matrix $S_{\ell_n X}$,
this summation can be easily carried out using the unitarity of 
the $S$ matrix
\begin{eqnarray}
\sum_{X \neq \ell_n} |S_{\ell_n X}|^2 = 1 -|S_{\ell_n \ell_n}|^2
\end{eqnarray}
Therefore, we only need to know the $S$ matrix elements of the elastic
scattering to determine the double differential cross sections for the
inelastic breakup, which can be written as    
\begin{eqnarray}
\frac{d^2\sigma(inelastic)}{d\Omega_p dE_p} & = & \rho(phase)
       \sum_{\ell_n m_n}(\sigma_{\ell_n}^{reaction}/\sigma_{\ell_n}^{elastic})
                              \nonumber \\
                & & \times \mid \beta_{\ell_n m_n}-\beta_{\ell_n m_n}^0 \mid^2
\end{eqnarray}
In this equation $\sigma_{\ell_n}^{reaction}$ and $\sigma_{\ell_n}^{elastic}$
are the reaction and elastic scattering cross sections for the neutron-target 
system corresponding to the partial wave $\ell_n$ respectively. 
$\beta_{\ell_n m_n}^{0}$ is defined in the same way as Eq. (17) with the
wave function $F_{\ell_n}$ being replaced by the
spherical Bessel function. More details of the derivation of the
inelastic breakup cross sections can be found in Refs.~\cite{Shy84,Pamp78}
 
The total cross section for the reaction (d,p) can
also be written as
\begin{eqnarray}
\sigma_{b-up(d,p)} & = & \frac{\pi}{k^2} \sum_{\ell} (2\ell+1)
                         T_{\ell}^{b-up(d,p)}. 
\end{eqnarray}
where $T_{\ell}^{b-up(d,p)}$ is the transmission coefficient for the 
(d,p) breakup reaction, which is also termed as the breakup probability
in ~\cite{Shy80}. The total breakup probability $T_{\ell}^{b-up,d}$
is given by 
\begin{eqnarray}
T_{\ell}^{b-up,d} & = & T_{\ell}^{b-up(d,pn)} + 
                          T_{\ell}^{b-up(d,p)}(inelastic)\nonumber \\ 
                  & &   + T_{\ell}^{b-up(d,n)}(inelastic)  
\end{eqnarray}
In Eq. (24), $T_\ell^{b-up(d,pn)}$, represents the breakup probability for 
the elastic breakup mode as defined above. The total breakup cross
section $\sigma_{b-up,d}$ is obtained
from $T_{\ell}^{b-up,d}$ by following an expression similar to Eq. (9).

\section{Results and discussion}

Apart from the zero range constant, finite range and
nonlocality parameters described already in section II, 
we require the optical potentials in the deuteron, proton and
neutron channels to calculate the breakup cross sections.
These have been taken
from the global sets given by Daehnick, Childs and Vrcelj~\cite{Dae80}
(for the deuteron channel)
and Becchetti and Greenlees~\cite{Becc69} (for the proton and 
neutron channels) respectively.

In Fig. 1 we show the results for the breakup probability for the
deuteron incident on a $^{51}$V target at the beam energy
of 56 MeV, calculated within the post form DWBA theory. In this
figure we have also shown 
the total transmission coefficients calculated with the same
deuteron optical potential.
We can see that the $(d,p)$ and $(d,n)$ breakup probabilities
are similar in shape and absolute magnitude. The elastic breakup
probability is much smaller and shows a different behavior as a function
of $\ell$. The cross sections for the inelastic (d,p), inelastic (d,n) and  
elastic (d,pn) breakup processes are 290 mb, 284 mb, and 122 mb
respectively, which lead to a total deuteron breakup cross section of 698 mb
for this target. The total (d,p) breakup cross section 
(which is the sum of inelastic (d,p) and elastic (d,pn) cross sections)
is 412 mb which is in reasonable agreement with the measured value of
481 mb ~\cite{Mat80}. The remaining difference between the experimental and
theoretical cross sections, to some extent, may be attributed to the likely 
contributions of quasi-elastic channels like (d,dp) to the
inclusive proton spectra. 
 
An important feature of Fig. 1 is the fact that for $\ell >$ 21, the 
elastic breakup probability (EBP) exceeds the total transmission 
coefficient. This shows the inadequacy of the phenomenological 
deuteron optical potential (used to determine the transmission
coefficient) for larger values of $\ell$. The 
Woods-Saxon shapes used to obtain such optical potentials 
are too restricted in their radius dependence. It is obvious that these
potentials do not account 
correctly for the elastic breakup channel (which we shall
refer to as dissociation in the following) at large distances.
We also note from this figure that both elastic and inelastic
breakup probabilities are relatively large at very small
partial waves. This is the consequence of weak absorption of neutrons
and/or protons. For breakup reactions involving heavier projectiles,
these probabilities are quite small at these partial
waves~\cite{Shy84,Shy80}.

\begin{figure}[here]
\begin{center}
\epsfxsize=8.3cm
\epsfbox{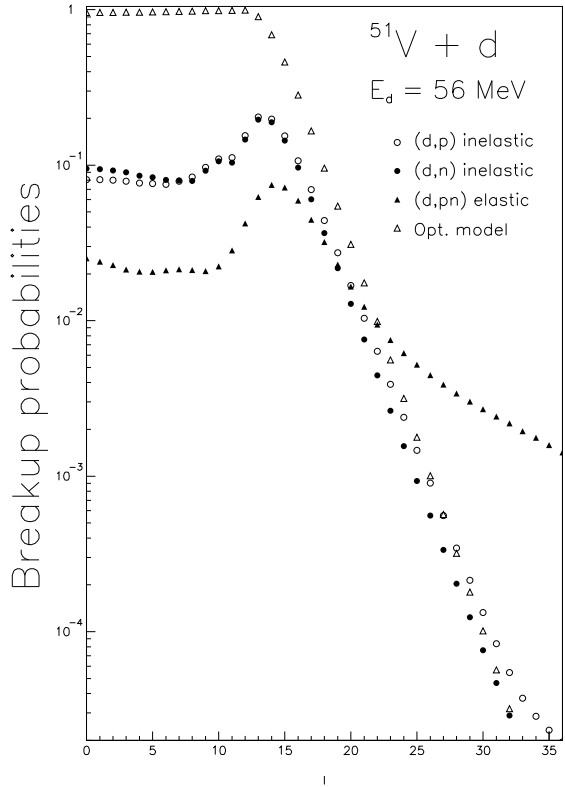}
\end{center}
\caption{
Calculated breakup probabilities in the scattering of 56 MeV deuteron
from $^{51}$V. The open triangles show the transmission coefficients
calculated with the optical potentials as explained in the text.}
\label{fig:figa}
\end{figure}

Now we proceed to determine the potential due to the dissociation process. 
As already pointed out in Ref.~\cite{Shy80}, it is the Coulomb
force which is 
responsible for the long range part of the EBP. 
In case of pure Coulomb interaction, the eikonal approximation may be
used to relate the imaginary phase shifts to an imaginary
potential~\cite{Sat90,Ing97}. Since $\sigma_{b-up,d}$ defines uniquely
an imaginary phase-shift, we can use this method to
determine the imaginary pure Coulomb dissociation potential by fitting
the EBP calculated by the post form DWBA theory with Coulomb
interactions only between deuteron and proton and the target nucleus.
 
On the left side of Fig. 2, the Coulomb EBPs are shown with    
Coulomb interactions obtained from: (1) an
uniform charge distribution (with a radius ($R_C$) of 1.3 fm)
(solid circles) and (2) a Woods-Saxon charge distribution with  
radius and diffuseness parameters of 1.3 fm and 0.65 fm respectively 
(open circles). We observe that the Coulomb dissociation is 
sensitive to the charge distribution  
for small $\ell$ values. For larger partial 
waves, however, they produce identical results which is to be expected.
It is explicitly clear (together with Fig. 1) that 
the Coulomb force determines the EBP at higher partial waves. We
have used EBP calculated with the Coulomb potentials 
generated from the Woods-Saxon charge distribution to perform the
fits. In fact, this diffuse Coulomb potential
has been used in all the post form DWBA calculations presented in
this paper.

\begin{figure}[here]
\begin{center}
\epsfxsize=8.30cm
\epsfbox{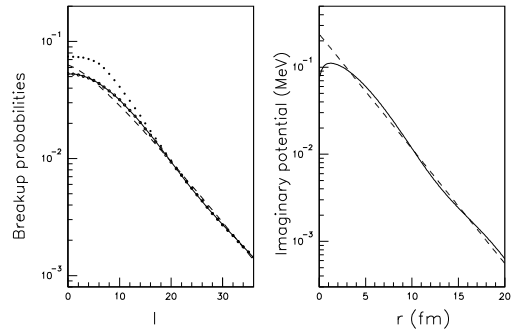}
\end{center}
\caption{
Results of calculations without nuclear interactions.
The left part of the figure shows the calculated Coulomb elastic 
breakup probabilities for the same reaction as in Fig. 1 obtained with
Coulomb potentials generated with a uniform charge distribution
(solid circles) and a charge distributions with Woods-Saxon shape
(open circles). The dashed and solid curves show the
best fits obtained by using 2 and 6 parameter parameterizations.
The right part shows the corresponding best
fit potentials obtained by fitting to the breakup probability
calculated with the Woods-Saxon charge distribution. The solid (dashed)
curve represents the results obtained with six (two) parameter 
parameterizations.} 
\label{fig:figb}
\end{figure}

We used a two parameter, ($ae^{-br}$), as well as a six parameter 
parameterization, $[(a_1+a_2 r^{1/2} + a_3 r + a_4 r^5)/(1+a_4 r)]
e^{(-b_1 r -b_2r^2)}$, in our fitting process. 
The best fit breakup probabilities obtained with
the two procedures are shown by dotted and full curves in the left part
of Fig. 2. The corresponding imaginary potentials are shown on the
right side of this figure. The six parameter search procedure 
(with $b_1$ and $b_2$ being equal to 0.1118 and 0.0113) provides
a better fit to the EBP. The two parameter fits had a 
radial dependence of $e^{-0.3r}$. In comparison to this  
the Woods-Saxon potential behave  approximately
as $e^{-r/a} \approx e^{-2.0r}$
(with $a$ being the diffuseness parameter),
at larger distances. Therefore, it is not surprising that the
transmission coefficients calculated with conventional optical
potentials drop too fast at larger partial waves. 

It is clear that potential due to the Coulomb
dissociation (PCD) is very weak. We found that the  
elastic scattering angular distributions calculated with 
phenomenological optical potential were almost unaffected by inclusion
of the PCD to its imaginary part. Also this leads to a change of 
less than 1\% in the total reaction cross section.   
The transmission coefficients at lower partial waves are affected
by the inclusion of PCD, but the nuclear effects damp them
out strongly in this region, as is discussed in the context of Fig. 3. 

We, therefore, added the PCD to the phenomenological imaginary
potential as it leads to the total transmission coefficient which
is larger than (or equal to the the EBP) for all the partial waves as
can be seen from the left part of Fig. 3, where the total transmission
coefficients obtained with (solid triangles) and without
(open triangles) adding the PCD to the phenomenological optical
potential are shown. It can be seen that, addition of PCD to the
phenomenological potential removes the anomaly described above.
In the right side of Fig. 3, we compare the results of the Coulomb EBP
obtained with (by using Eq. (13)) and without nuclear distortion.
We note that nuclear distortion effects suppress the Coulomb 
EBP strongly for lower partial waves.  However, for larger
partial waves the two calculations produce identical
results. 
 
\begin{figure}[here]
\begin{center}
\epsfxsize=8.30cm
\epsfbox{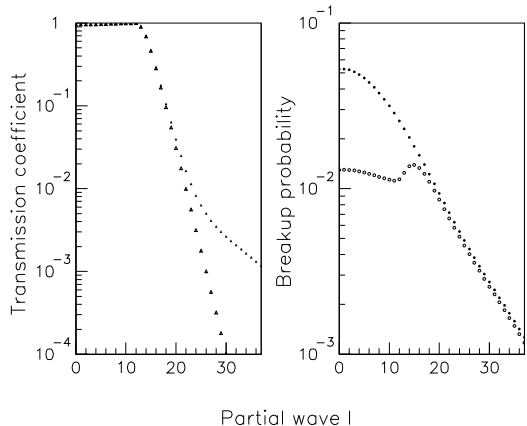}
\end{center}
\caption{ 
The left part shows the transmission coefficients in the elastic
scattering calculated with the phenomenological optical potential with
(solid triangles) and without (open triangles) addition of the 
imaginary potential due to the elastic Coulomb
dissociation. The right part shows the breakup probabilities for elastic
Coulomb dissociation calculated in the eikonal approximation (same curve as
shown in figure 2) (solid circles) and those obtained in the full
calculations with nuclear distortions included (open circles).}
\label{fig:figc}
\end{figure}

In order to find the potentials due to the inelastic breakup
and nuclear dissociation,
we used Eq. (13). The inelastic and elastic breakup probabilities
calculated within post form DWBA (Fig. 1)  were fitted by using 
a sum of the Gaussians (SOG) parameterization~\cite{Devr87}
\begin{eqnarray}
U_{bu}^{SOG}(r) & = & \sum_i A_i[exp(-\{(r-R_i)/\gamma\}^2) \nonumber \\
                          & &      +exp(-\{(r+R_i)/\gamma\}^2)]
\end{eqnarray}
The value of $\gamma$ was taken to be $\sqrt{2/3}$~\cite{Devr87},
and a total of 15 terms were included in the sum.
The coefficients
$A_i$ and positions $R_i$ were varied in order to get the 
best fits to the elastic, inelastic and total breakup probabilities,
which are shown,
in the left, middle and right parts of Fig. 4 respectively.
The solid circles show the results of post form DWBA (same as that in
Fig. 1), while open circles represent our fits. In case of EBP,
we also show the results for pure Coulomb case (open triangles)
(the same as shown in the right side of Fig. 2).
We can see that Coulomb EBP agree very
well with that obtained by fitting the total EBP for values of $\ell$ $>$
21. It should be remarked here that in the study of 
Christley et al~\cite{Chr91}, who have used a similar method to determine
the fusion potential by fitting to the fusion cross sections calculated
within a coupled-reaction-channel model, the quality of fits were 
not as good. They attribute this failure to the lack of
$\ell$-dependence in their fitting potential (which were taken to be of
the Woods-Saxon type). However, we do not require such a
dependence in our fitting procedure.

\begin{figure}[here]
\begin{center}
\epsfxsize=8.30cm
\epsfbox{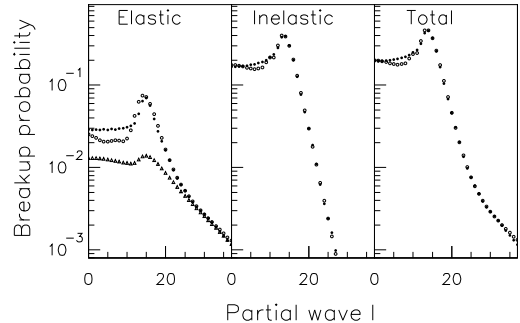}
\end{center}
\caption{
The elastic, inelastic and total breakup probabilities 
calculated within the post form distorted-wave Born-approximation
(solid circles) and our best fit values to them (open circles). Open
triangles, in the left part of the figure, represent
the breakup probability for the elastic 
Coulomb dissociation (same as those shown in the right part of
fig. 3 by open circles).}
\label{fig:figd}
\end{figure}

The potentials giving these breakup-probabilities are shown in the upper
part of Fig. 5. The solid, dashed and dotted curves show the potentials
due to the inelastic breakup (which has been plotted after multiplying the
actual values by 0.09), nuclear dissociation and Coulomb dissociation
(already shown in Fig.2) respectively. We note that the inelastic breakup
and nuclear dissociation potentials are strongly concentrated in a  
region around 6.5 fm and 8 fm, respectively. Both potentials have about
the same width, 2 fm (FWHM), which is very similar to the rms-radius of
the  deuteron. The difference in the localization of the inelastic
and dissociation potentials confirms the fact that inelastic breakup
of the projectile occurs in regions closer
to the target nucleus as compared to the elastic breakup.
It is worthwhile to note that the potential due to the
inelastic breakup is quite large. Therefore, those 
calculations~\cite{Sak86,Yab92,Baur76,Ber89} where this mode of
breakup is not
considered are likely to produce weaker imaginary  potentials
due to the breakup channels as compared to ours.
Due to its long range, the potential due to the Coulomb dissociation is
still of appreciable magnitude at large distances.  Coulomb
breakup thus, can  take place even outside the charge distribution 
of the target nucleus.   

In the middle part of Fig.5, we compare the imaginary part of the 
phenomenological optical potential (solid line) with the  bare
potential (obtained by subtracting the potential due to breakup from it)
(dashed line). It is clear that breakup is the dominant   
absorption effect in the surface region. The bare potential
could be the starting point of double folding model calculations.

\begin{figure}[here]
\begin{center}
\epsfxsize=8.30cm
\epsfbox{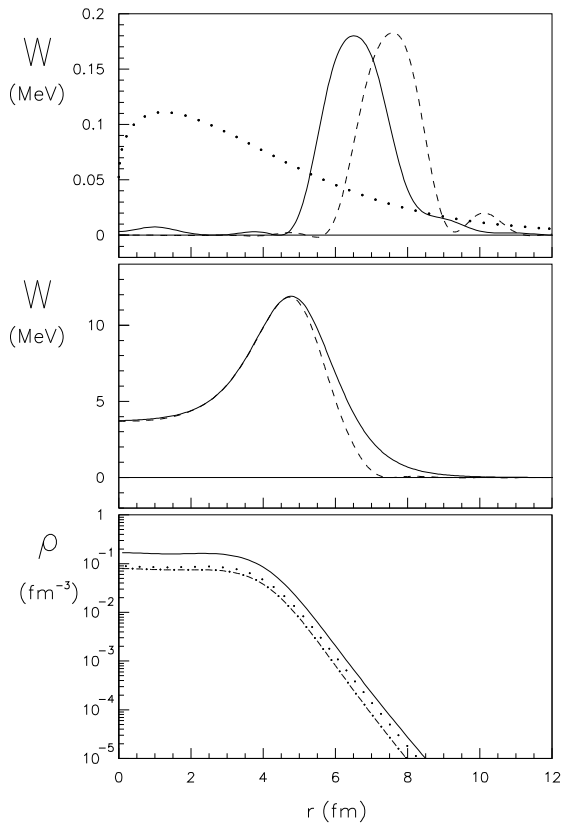}
\end{center}
\caption{ 
The upper figure shows the imaginary potentials due to inelastic
breakup (arbitrarily multiplied by 0.09 to fit in the plot) (solid line)
and the potential due to nuclear dissociation (dashed line).
The dotted curve shows the Coulomb dissociation potential.
The middle part shows the optical potential with (solid curve) and without 
(dotted curve) effects of breakup.
In the lower part
the dashed, dotted and solid curves show the neutron,  proton
and matter densities for $^{51}V$ respectively, as  calculated by
Fayans~\protect\cite{Fay98}.}
\label{fig:fige}
\end{figure}

In the lower part of Fig. 5, we show the radial dependence of the
neutron, proton and matter densities for the $^{51}$V nucleus~\cite{Fay98}.
WE note that breakup probabilities peak in a region where the 
matter and charge densities are much smaller as compared to the
central density of the target nucleus. Therefore, breakup  
process is really sensitive to the extreme peripheral regions. This 
also confirms the spectator role of the target nucleus in the 
breakup process (Serber picture). However, the region where the
breakup potentials peak is much larger than the half density radius
deduced by Serber~\cite{Ser47}. 

It should be stressed that our method of obtaining the imaginary part
of the optical potential due to the breakup channels is 
phenomenological in nature. However, it has several good features. The
distinctly different nature of the total breakup cross section as a 
function of the incident partial waves is automatically taken into
account. The Coulomb breakup process which is responsible for the long
range part of the elastic breakup probability (and the dissociation 
potential) is included into our calculations. Furthermore, we include
the inelastic breakup mode which makes the largest contribution to the
imaginary potential. Both these effects were ignored in  
calculations presented in Refs.~\cite{Sak86,Yab92}. 

\section{Summary and conclusions}

In this paper we calculated the imaginary part of the optical
potential due to the breakup channels by using a phenomenological method
in which contributions to the reaction cross sections from the elastic
and inelastic breakup cross sections of the projectile in the field of
target nuclei, calculated in the post form distorted-wave
Born-approximation, are used as constraints in a fitting procedure. 
In this method, only the first order breakup process (the coupling of
the elastic channel to the breakup channel) is considered. Then 
unitarity of the $S$-matrix is used to determine the influence of
breakup on the elastic channel.  

The Coulomb part of the imaginary dissociation potential (obtained from 
the pure Coulomb elastic breakup probabilities) accounts for
the long range part of the absorption and removes the anomaly 
where for the partial waves beyond 20, the elastic breakup probabilities
were found to be even larger than the transmission coefficients
calculated with the usual phenomenological optical potentials.
To our knowledge this is the first calculation where the absorption
due to the Coulomb dissociation process has been included in the 
optical model. The potentials due to the nuclear dissociation
are found to be peaked in the region around 8.0 fm.     

The imaginary potential due to inelastic breakup is about an order of
magnitude larger than that due to the dissociation process, and is
concentrated at somewhat shorter distance (around 6.5 fm) as 
compared to the latter. This suggests that the inelastic breakup takes
place at distances closer to the target nucleus in comparison to the 
elastic breakup. The magnitude of the total imaginary potential,
due to the breakup channels (IPBC) is, therefore, almost
solely due to the inelastic breakup process in the region around 6-8 fm.
This is consistent with the dominant contribution of this mode 
to the total breakup cross section. This is the main reason 
for our IPBC  being stronger than the imaginary
part of the dynamical polarization potential due to breakup channels
calculated by other authors, who have ignored the inelastic breakup
mode.  
 
Our method can be applied to any projectile having a strong breakup
channel. Therefore, It would be interesting to use this method to
the elastic scattering of halo nuclei (eg. $^{11}$Li
and $^{11}$Be) ~\cite{Kol92}, where the breakup
cross sections are significantly enhanced and the effect of breakup
on the elastic scattering is very strong.

This work has been supported by the Wenner-Gren Center Foundation,
Stockholm. One of the authors (RS) would like to acknowledge several
useful discussions with Prof. I.J. Thompson of University of 
Surrey, Guildford. We also want to thank Sergei Fayans for providing
the densities for $^{51}$V.

\end{document}